\documentclass[nobibnotes,amssymb,aps,twocolumn,superscriptaddress,floatfix,longbibliography]{revtex4-2}
\usepackage[dvips]{graphicx}
\usepackage{dcolumn}
\usepackage{bm}
\usepackage{amsmath}
\usepackage{amssymb}
\usepackage{ulem}
\usepackage{setspace}
\usepackage{xcolor}
\usepackage{float}
\graphicspath{{./figs/}}
\usepackage{empheq}
\usepackage{lipsum}

\makeatletter
\let\SIrestore@title\title
\let\SIrestore@author\author
\let\SIrestore@affiliation\affiliation
\let\SIrestore@thanks\thanks
\let\SIrestore@date\date
\let\SIrestore@maketitle\maketitle
\let\SIrestore@and\and
\makeatother

\usepackage[compact]{titlesec}
\titlespacing{\section}{10pt}{*2}{*2}
\titlespacing{\subsection}{10pt}{*2}{*2}
\titlespacing{\subsubsection}{10pt}{*4}{*4}

\newcounter{boxlblcounter}  

\newcommand{\ddo}{D$_2$O }
\newcommand{\ddox}{D$_2$O}
\newcommand{\ddocoin}{D$^+$/D$^+$/O$^+$ }
\newcommand{\ddocoinx}{D$^+$/D$^+$/O$^+$}
\newcommand{\dodcoin}{D$^+$/OD$^+$ }
\newcommand{\dodcoinx}{D$^+$/OD$^+$}

\newcommand{\StanfordPhysics}{
Department of Physics, Stanford University, Stanford, CA 94305, USA}
\newcommand{\StanfordAP}{Department of Applied Physics, Stanford University, Stanford, CA 94305, USA}
\newcommand{\PULSE}{
Stanford PULSE Institute, SLAC National Accelerator Laboratory, Menlo Park, CA 94025, USA}
\newcommand{\SLAC}{Linac Coherent Light Source, SLAC National Accelerator Laboratory, Menlo Park, CA 94025, USA}
\newcommand{\Davis}{Department of Chemistry, University of California, Davis, One Shields Avenue, Davis CA 95616}

\begin{document}
\title{Electron spectra from strong-field enhanced ionization in heavy water}
\author{Eleanor Weckwerth}
\address{\PULSE}
\address{\StanfordPhysics}

\author{Chuan Cheng}
\address{\PULSE}
\address{\StanfordPhysics}

\author{Ian Gabalski}
\address{\PULSE}
\address{\StanfordAP}
\address{\Davis}

\author{Andrew~J.~Howard}
\address{\PULSE}
\address{\StanfordAP}
\address{\Davis}

\author{Mathew~Britton}
\address{\PULSE}
\address{\SLAC}

\author{Aaron~M.~Ghrist}
\address{\PULSE}
\address{\StanfordAP}

\author{Haoran Ma}
\address{\PULSE}
\address{\StanfordPhysics}

\author{Salma A. Mohideen}
\address{\PULSE}
\address{\StanfordPhysics}

\author{Philip H. Bucksbaum}
\address{\PULSE}
\address{\StanfordPhysics}
\address{\StanfordAP}
\thanks{Corresponding author: phbuck@stanford.edu}

\date{\today}

\begin{abstract}
Strong-field enhanced ionization (EI) is a phenomenon in which stretching of interatomic bonds into a distorted molecular geometry leads to an increase in the tunneling ionization rate driven by a strong field.
While the nuclear dynamics leading to EI are well established, the electron dynamics of the ionization step remain largely unexplored. 
Isolating the momentum distribution of the electrons involved in EI is critical to fully characterizing the phenomenon. 
We have measured the electron momentum distribution associated with EI in triple ionization of \ddo 
using 6-fs pulse pairs together with full fragment momentum imaging and electron-ion correlation methods.
We find that the EI electron momentum distribution differs substantially from that of
standard strong-field tunneling from molecules, exhibiting an increased yield of electrons with large momentum in the direction of the laser polarization, and a change from the expected Gaussian distribution.
These observations indicate that the instantaneous EI tunneling rate is maximized at a critical value of the laser electric field, rather than at the peak of an optical cycle.  This finding distinguishes EI from Keldysh tunneling rate predictions, where tunneling rate increases monotonically with field strength.
These pronounced differences between EI and non-EI electron spectra are critical tests of models of enhanced ionization and 
suggest a route towards control of the sub-cycle timing of electron emission. 

\end{abstract}

\maketitle



Strong laser fields can induce tunnel ionization in atoms and molecules through 
a quasi-static barrier in the 
combined potential of the ion and the laser field, 
and thus provide a powerful way to induce electron dynamics in Coulomb-bound quantum systems \cite{Keldysh1964_Ionization, Ammosov1986_Tunnelb, Perelomov1966_Ionization, Corkum1993_Plasmaa, Krause1992_Highorder}. 
Atomic motion initiated by photoexcitation or photoionization of molecules 
leads to a range of related strong-field phenomena \cite{Bucksbaum1990_Softening, Frasinski1987_Femtosecond, Ibrahim2018_H2}. 
One example 
is strong-field enhanced molecular ionization (EI), where the ionization probability increases dramatically when the molecule reaches a critical geometry during its nuclear evolution \cite{Posthumus2004_dynamics}. 
EI has been studied in diatomic molecules \cite{Zuo1995_Chargeresonanceenhanced, Posthumus1995_Dissociative, Seideman1995_Rolea, Trump1999_Multiphotona, Trump1999_Strongfield, Ben-Itzhak2008_Elusive, Xu2015_Experimental, Ergler2005_TimeResolveda, Legare2003_TimeResolved} and in more complex systems, chiefly by analyzing Coulomb explosions following multiple ionization \cite{Howard2023_Filminga, Howard2024_Uncovering, Howard2025_Isotopeselective, McCracken2017_Geometrica, Gong2014_StrongField, Erattupuzha2017_Enhanced, Burger2018_Timeresolved, Mohideen2025_Orientationdependent,  Bocharova2011_Charge, Brichta2006_Ultrafast, Song2022_Dissociative}. 

In the standard description of strong-field EI, 
rapid nuclear motion 
brings the system to a critical geometry 
in which multiple atoms are aligned along the laser polarization and the tunneling barrier is reduced. 
The field distorts the molecular electrostatic potential (MEP), trapping the electron density on an uphill atom, away from the downhill Coulomb saddle point \cite{Zuo1995_Chargeresonanceenhanced, Wu2012_Probing}. 
In diatomic molecules, the bond stretches 
and the potential barrier between the atoms is lowered. 
At some critical separation, the 
barrier suppression is sufficient to release the trapped electrons, allowing them to tunnel directly into the continuum at energies above the Coulomb saddle point.
Specifics can vary in more complex molecules, but the common EI mechanism is that nuclear motion leads to transient molecular geometries where the tunneling barrier is significantly reduced and field ionization is consequently enhanced. 


Water, including heavy water (\ddox), exhibits EI of the triple ionization channel leading to \ddocoin Coulomb explosion. Strong-field double ionization initiates unbending and symmetric bond stretching in the dication, leading to a ``slingshot'' motion \cite{Howard2023_Filminga, Howard2024_Uncovering, Howard2025_Isotopeselective}. About 20~fs after double ionization, the molecule reaches a critical stretched linear geometry where the probability of further strong-field ionization increases sharply, consistent with EI predictions. In this geometry, an internal barrier between the oxygen and the downhill deuteron traps electron density on the oxygen. Ionization then proceeds through tunneling through this internal barrier, producing a substantially narrower effective tunneling barrier and a corresponding increase in the ionization rate. 

Most prior studies of strong-field EI have focused on the ionic fragments from Coulomb explosion, providing insight into the nuclear dynamics leading to the critical geometry but only indirect access to the ionization process itself. 
Field-ionized electron momentum distributions can provide direct information about the dynamics of EI, probing how the evolving molecular structure modifies the strong-field ionization process beyond simply increasing the ionization probability.
However, measuring this requires difficult high-order coincidence methods to identify the electrons that coincide with the 
Coulomb explosion of a highly charged final state.

Here we report our measurements of the electron momentum distribution associated with EI in heavy water using 6-fs pulse pairs and electron-ion correlation methods. We find a strong departure from the familiar Keldysh tunneling, where the ionization rate increases monotonically with field strength during an optical cycle. Instead, we find the maximum tunneling rate occurs at a critical field strength that may not be the field maximum, which broadens and shifts the ionized electron momentum distribution. 


\begin{figure}[htbp]
	\centering
	\includegraphics[width=0.9\linewidth]{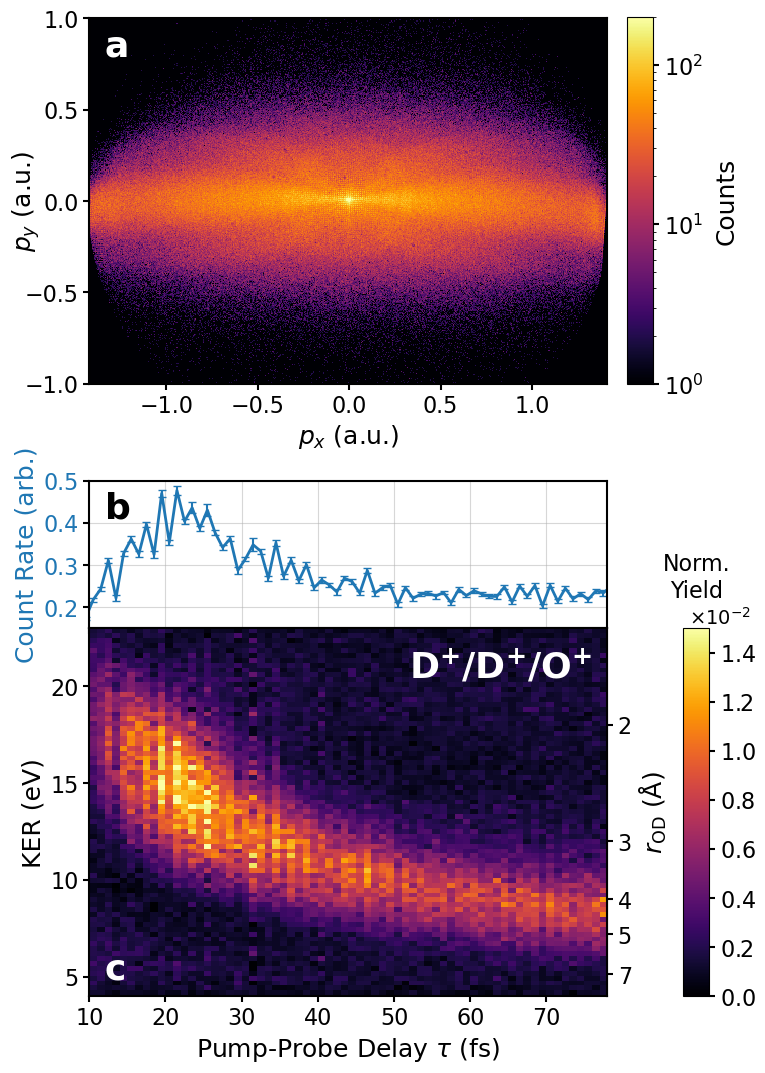}
	\caption{Experimental photoelectron and ion momentum observables. 
    \textbf{(a)}~Photoelectron momentum distribution (PMD) produced from a single (probe) pulse polarized along the detector $x$ axis. The longitudinal momentum $p_x$ reflects the electron momentum along the laser polarization axis and encodes the tunneling dynamics at the time of ionization.
    \textbf{(b)}~Normalized yield of the \ddocoin triple ionization channel as a function of interpulse delay. The yield peaks near 20~fs corresponding to the time it takes for the dication to reach the critical geometry for EI. \textbf{(c)}~Kinetic Energy Release (KER)-resolved yield of the \ddocoin triple ionization channel over the same delay range as in panel (b). The data show a characteristic Coulomb explosion curve caused by the dissociation and stretching of the molecule before the probe pulse arrives to ionize the rest of the way to the \ddox$^{3+}$ charge state.
    } 
	\label{fig:exp_overview}
\end{figure}

Gaseous \ddo is ionized by a pair of cross-polarized 6-fs few-cycle pulses (central wavelength 800~nm) with variable delay and peak electric field strengths in the range of 0.13-0.19 a.u. (corresponding to intensities in the range of $6 \times 10^{14}~\mathrm{W/cm^2}$ to $1.3 \times 10^{15}~\mathrm{W/cm^2}$). Further details are listed in the Supplemental Materials.
The resulting electrons and ion fragments are detected in coincidence using a dual electron and ion momentum spectrometer, which detects the full vector momentum of each charged ion fragment and two momentum components of each electron projected into the plane perpendicular to the spectrometer axis \cite{Zhao2017_Coincidencec,Zhao2017_Coincidence, Gabalski2026_Fasta}.
The pump pulse is polarized along the time-of-flight $\hat{z}$-axis of the spectrometer, while the probe pulse is polarized in the detector plane $x$-axis, so that electron momenta parallel and perpendicular to the probe laser polarization can both be measured. 
The triple ionization events are identified by requiring momentum-conserving coincidence of the three ion fragments of the associated trication Coulomb explosion \ddox$^{3+} \rightarrow$ \ddocoinx. 


A typical photoelectron momentum distribution (PMD) ($p_x,p_y$) integrated over $p_z$ is shown in Fig.~\ref{fig:exp_overview}a for a single (probe) 6-fs pulse polarized along $x$. The $p_x$ distribution reflects the sub-cycle distribution of tunneling times because the electron acquires additional drift momentum in this direction approximately equal to the vector potential at the moment of ionization. The $p_y$ and $p_z$ components display cylindrical symmetry, so the PMD could be converted to ($p_x,p_\perp$) by an inverse Abel transform; however, the cross-polarized pump-probe measurements break cylindrical symmetry and so cannot be Abel transformed. Therefore all PMDs are shown as raw projections for consistency. 
The familiar above-threshold ionization (ATI) rings due to multicycle interference are not evident here because the 6-fs pulse consists of only about two optical cycles \cite{Wickenhauser2006_Laserinduced, Brabec2000_Intense, Freeman1987_Abovethresholda}.
Some holographic features of strong-field tunneling are seen, such as the spider-leg pattern \cite{Werby2021_Dissecting, Huismans2011_TimeResolved, Hickstein2012_Directb, Gong2016_Pathwayresolved, Moller2014_Offaxisa}. The distribution cuts off at 1.4 a.u. due to limitations of the voltage-switching detector, which prevents simultaneous detection of higher-momentum electrons and ions.






To isolate electrons associated with \ddocoin production, the photoelectron momentum distribution correlated with a given ion coincidence channel is extracted using covariance mapping  \cite{Cheng2023_Multiparticlea, Allum2022_localized, Cheng2026_Imaging}, through evaluation of $\mathrm{Cov}[e^{-}(p_x), N_i]$, where $e^{-}(p_x)$ is the electron longitudinal momentum and $N_i$ is the ion coincidence channel yield. The covariance-selected spectrum contains electrons emitted by both pulses as no background subtraction between pump and probe electron contributions is performed. Their contributions are instead distinguished by the orthogonal pulse polarizations and the resulting strongly localized momentum distribution of the pump electrons, as discussed below and in the Supplemental Material. 
The yield of the \ddocoin triple ionization channel versus interpulse delay is shown in Fig.~\ref{fig:exp_overview}b. Ionization enhancement is evident for pulse separations of approximately 20~fs, corresponding to the time required for the dication to evolve along the slingshot trajectory to the critical geometry for EI \cite{Howard2023_Filminga}. 
When the pump intensity is held fixed and the probe intensity is varied, the ratio of the peak yield near 20 fs and the yield at late delays remains approximately constant, even though higher intensity would generally be expected to increase the production of higher charge states (see Supplemental Material for more details). 
The energy-resolved plot in Fig.~\ref{fig:exp_overview}c demonstrates the geometric evolution of the molecule over time, and the  
approximate O-D separation is inferred from the ``Coulomb explosion'' energy of dissociation.
Maximum enhancement occurs for 
O-D bond lengths of approximately 2.5 \AA, consistent with previous measurements of EI in \ddo \cite{Howard2025_Isotopeselective}.

\begin{figure}[htbp]
	\centering
	\includegraphics[width=0.9\linewidth]{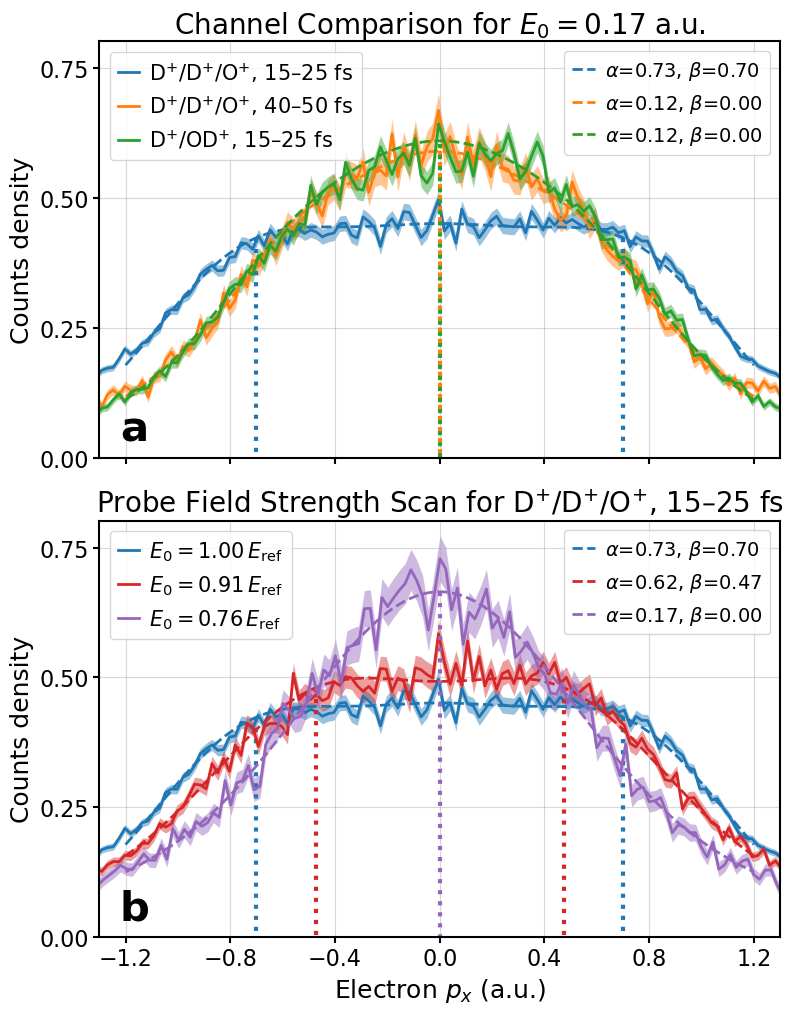}
	\caption{
    Electron $p_x$ distribution in covariance with triple and double ionization ion coincidence events. In both panels, the shaded region indicates the estimated statistical uncertainty for each channel, dashed lines show fits to Eq.~\ref{eq:gaussian_sum}, and resulting values of relative contribution factor $\alpha  = \left(A_1\right)/\left(A_0 + A_1\right)$ and primary offset $\beta$ are shown in the upper right legend. Vertical dotted lines display the value of $\pm \beta$ for each curve.
    \textbf{(a)}~Comparison of electron $p_x$ spectrum for fixed probe field strength $E_0 = 0.17$ a.u. The \ddocoin triple ionization channel at delays corresponding to the EI geometry (blue) compared to the same channel at larger interpulse delays (orange) and the \dodcoin double ionization channel over the same delay range (green). 
    \textbf{(b)}~Electron $p_x$ distribution for the EI \ddocoin channel at varying probe field strengths relative to the reference field $E_{\mathrm{ref}} =$~0.17~a.u., including $1.00\,E_{\mathrm{ref}}$~(blue), $0.91\,E_{\mathrm{ref}}$~(red), and $0.76\,E_{\mathrm{ref}}$~(purple). 
    }
	\label{fig:covariance_1d}
\end{figure}

The EI study is performed with a pump pulse energy that produces dications but is not intense enough to triply ionize at any significant rate, with the single pulse \ddocoin coincidence rate less than 2\% of the rate measured at late delays when the pulses are well separated. Therefore the third electron in the \ddocoin channel is overwhelmingly emitted by the probe pulse, which is polarized along the detector $x$-axis, while electrons from the orthogonally polarized pump pulse are concentrated near $p_x \approx 0$. Combined with the reduced detection efficiency at low momentum, this means the measured $p_x$ distribution is dominated by probe-ionized electrons, so although the covariance-selected electron spectrum contains contributions from both pulses, it primarily reflects the dynamics of the final ionization step. More details of the pump electron contribution are listed in the Supplemental Material.

The measured electron $p_x$ distributions in covariance with the triple ionization channel (\ddocoinx) and a representative double ionization channel (\dodcoinx) are shown in Fig.~\ref{fig:covariance_1d}a, with shaded bands indicating the statistical uncertainty from error propagation. Each spectrum is integrated over a 10-fs delay window. For the triple ionization channel, one window is chosen within the EI region identified in Fig.~\ref{fig:exp_overview}b (EI \ddocoin spectrum), and a second window of equal duration is taken outside this region (non-EI \ddocoin spectrum). For comparison, a double ionization spectrum is constructed over the same EI delay window (non-EI \dodcoin spectrum). The EI \ddocoin spectrum differs from both the non-EI \ddocoin spectrum and the non-EI \dodcoin spectrum, indicating distinct electron dynamics associated with the enhanced ionization regime.
The EI \ddocoin spectrum is noticeably broader, with a higher proportion of counts at large momentum and a significantly different shape than the other two distributions. As shown in Fig.~\ref{fig:covariance_1d}b, this broadening decreases as the probe field strength is reduced.

The PMD predicted by strong-field Keldysh ionization is a Gaussian distribution centered on $p_\parallel=0$, with a width that scales as the laser vector potential $A$ \cite{Keldysh1964_Ionization,Ammosov1986_Tunnelb,Perelomov1966_Ionization,Corkum1993_Plasmaa}. (See Supplemental Information for a brief derivation.) 
This is the well-established Strong-Field Approximation (SFA), which maps the value of $p_\parallel$ to the field phase at the moment of tunnel ionization \cite{Corkum1993_Plasmaa}. The highest tunneling rate occurs at the peak field, which yields electrons with almost no drift momentum, i.e. $p_\parallel=0$. Electrons with nonzero values of $p_\parallel$ have tunnel ionized at earlier or later phases in the optical cycle. 
The increased width of the EI electron distribution therefore indicates that for this channel, ionization occurs over a wider range of times within the cycle. 
As the probe field strength is reduced, the distribution narrows, indicating a smaller spread of ionization times within the cycle.

Deviations from this simple Gaussian shape from SFA accompany some important discoveries about strong-field processes.  
For example, electron-ion rescattering changes the high energy tail of the PMD, leading to high harmonic generation (HHG). Ponderomotive gradient forces shift the distribution when laser pulse durations exceed a few picoseconds. Circular polarization alters tunneling, as do internal dynamics such as Freeman resonances and shake-up \cite{Hofmann2014_Interpreting,Freeman1987_Abovethresholda,Wolter2014_Formation}; however, none of these pertain to our experimental conditions. 
The deviations observed here therefore reflect a distinct modification of the ionization dynamics associated with strong-field EI.

The EI distributions exhibit maximum yield at a momentum displaced from zero, in contrast to the zero-centered Gaussian distribution for Keldysh tunneling discussed above. We therefore use an empirical displaced Gaussian model to characterize this behavior consistently across datasets. To quantify the measured deviation from a zero-centered Gaussian, each $p_x$ spectrum is fit to a model consisting of a Gaussian centered at zero together with a symmetric pair of offset Gaussians centered at $\pm \mu_1$,
\begin{equation} \label{eq:gaussian_sum}
\begin{split}
f(p_x)  & = A_0 \left[G\left( p_x, 0, \sigma_0\right)\right] \\ 
&\quad + \frac{A_1}{2} \left[G\left( p_x, \mu_1, \sigma_1\right) + G\left( p_x, -\mu_1, \sigma_1\right)\right]
\end{split}
\end{equation}
where $G(x,\mu,\sigma) = \exp\!\left[-(x-\mu)^2/(2\sigma^2)\right]$ is a Gaussian centered at $\mu$ with width $\sigma$.
The center of the distribution where the pump electrons are concentrated is excluded from the fit, minimizing the pump electron contribution so that the fit primarily reflects the probe electron distribution. 
Each spectrum is characterized by the relative weight of the offset component,
$\alpha  = \left(A_1\right)/\left(A_0 + A_1\right)$,
and by the position of the dominant Gaussian component, which we define as a derived quantity $\beta$, where $\beta = 0$ when $\alpha \le 0.5$ and $\beta = \mu_1$ otherwise,
so that $\beta = 0$ corresponds to a distribution dominated by the centered Gaussian, while $\beta = \mu_1$ indicates dominance of the offset Gaussian pair. 
While this empirical model does not fully reproduce the detailed structure of the measured distributions, it provides a simple framework for quantifying the departure from the centered Gaussian and the displacement of the enhanced yield without imposing a more complex model. 
For both the non-EI \ddocoin and non-EI \dodcoin spectra, the fitted distributions are dominated by the central Gaussian component ($\alpha < 0.5$), indicating that the spectra are well described by a single Gaussian. 
However, the EI \ddocoin spectrum at the highest field strength $E_{\mathrm{ref}}$ shows a dominant offset component, with $\alpha = 0.73$, indicating that the distribution is dominated by the offset component. In the field strength scan in Fig.~\ref{fig:covariance_1d}b, the values of both $\alpha$ and $\beta$ decrease as the probe field is lowered, indicating a reduced contribution of the offset component and a smaller displacement of the peak from zero.

\begin{figure}[htbp]
	\centering
	\includegraphics[width=0.9\linewidth]{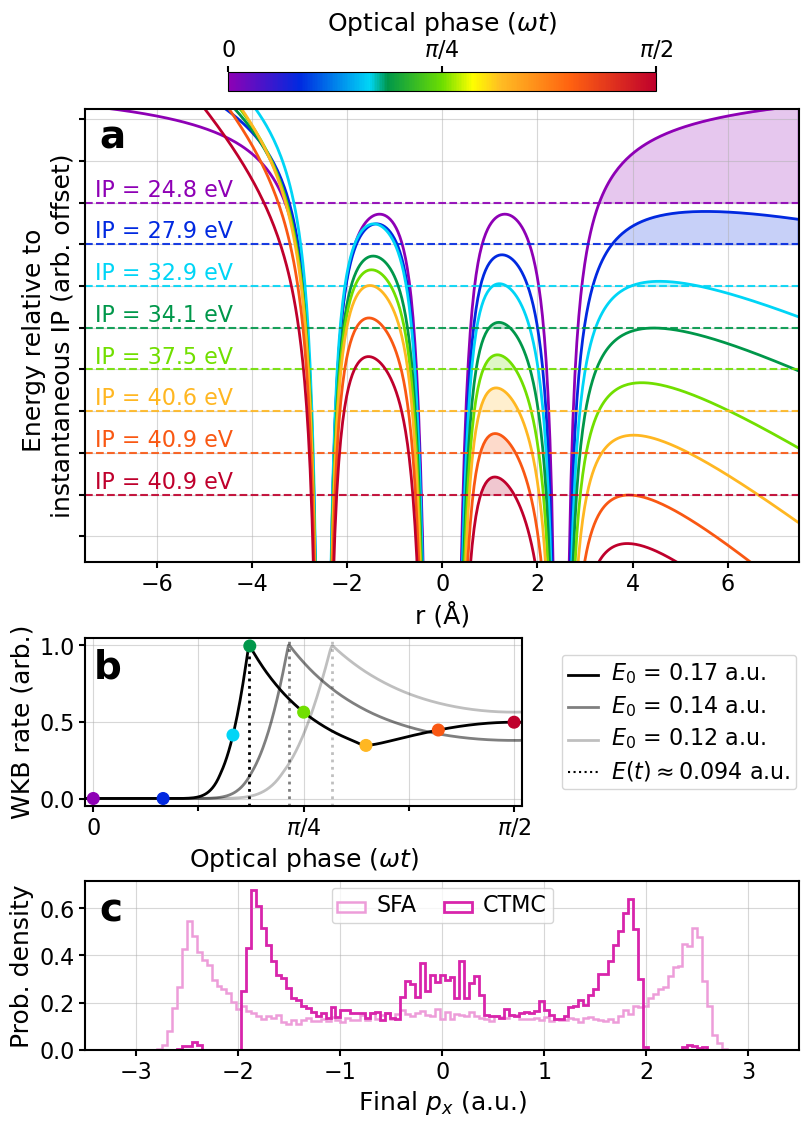}
	\caption{One-dimensional WKB calculations of tunneling rates from linear \ddo as a function of optical phase.
    \textbf{(a)}~Molecular electrostatic potential (MEP) of linear stretched \ddox$^{3+}$ ($r_{\mathrm{OD}} = 2.50$ \AA) calculated 
    for several instantaneous field strengths between $\phi=\omega t = 0$ and $\pi/2$ in a sinusoidal field with peak strength 0.17 a.u. For each phase, the MEP is plotted relative to the instantaneous \ddox$^{2+}$ ionization potential (IP) shown as a dashed line. 
    All curves are shown on the same energy scale but vertically offset for clarity, with the offsets chosen such that the instantaneous IP at successive phases are evenly spaced. The downhill tunneling barrier is shaded, indicating the classically forbidden region used in the WKB calculation.
    \textbf{(b)}~WKB tunneling rate as a function of optical phase for tunneling through the combined internal and external barrier, shown for three different peak field strengths $E_0$. Points along the black line indicate the phases corresponding to the curves shown in panel~(a). The field-dependent tunneling rate is the same for all three curves, but maps onto optical phase differently for each peak field strength. Vertical dotted lines indicate the phase at which the instantaneous field reaches 0.094 a.u. In each case, the peak tunneling rate occurs at this critical field strength, before the peak of the field. 
    \textbf{(c)}~Final longitudinal momentum distributions calculated for $E_0=0.17$ a.u. using the SFA and a one-dimensional semiclassical trajectory Monte Carlo (CTMC) model. Ionization times and tunnel-exit positions are sampled from the WKB calculation in panel (b). The SFA maps each ionization time directly to $p_x=-A(t_0)$, while the CTMC calculation propagates the outgoing electron in the laser field and molecular potential.
    }
	\label{fig:MEP_fig}
\end{figure}


Fig.~\ref{fig:MEP_fig} shows a simple model to explain these observations. 
The MEP of model linear stretched \ddox$^{3+}$ at the critical geometry identified in Ref.~\cite{Howard2025_Isotopeselective} is shown in Fig.~\ref{fig:MEP_fig}a for several optical phases $\phi = \omega t$ spanning $0$ to $\pi/2$ in a laser cycle with peak field $E_0 = 0.17$~a.u. The corresponding instantaneous ionization potential (IP) of \ddox$^{2+}$ at each phase is indicated by a horizontal dashed line. The curves are plotted on a common energy scale but vertically offset for clarity, and the downhill portion of the tunneling barrier is shaded to indicate the classically forbidden region relevant for tunneling.
When the field is zero at $\phi=0$, the electron is confined by an effectively infinitely wide barrier, with the IP lying above the internal barrier. As the field turns on, the instantaneous IP becomes more negative due to the DC Stark shift, while the MEP is simultaneously modified by the presence of the field such that both the internal and external barriers are suppressed. The system moves through a region where the width of the internal barrier is maximally suppressed, shown in Fig.~\ref{fig:MEP_fig}a by the dark green curve. Beyond this point, as the field continues to increase, the IP is driven further downward while the internal barrier is not substantially further reduced, so the effective barrier width and height increase relative to earlier phases. 

The WKB tunneling rate through the classically forbidden barrier region of \ddox$^{2+}$ is shown in Fig.~\ref{fig:MEP_fig}b \cite{Wentzel1926_Verallgemeinerung, Kramers1926_Wellenmechanik, Brillouin1926_mecanique, Keldysh1964_Ionization}. 
The ionization rate is shown for three peak field strengths ($E_0 = 0.17$, $0.14$, or $0.12$ a.u.) to emphasize that the maximum occurs at an instantaneous field strength of approximately 0.094 a.u. (dark green curve in Fig.~\ref{fig:MEP_fig}a), before the peak of the field. 
This is due to the interplay between the instantaneous DC Stark shift and the interatomic barrier suppression, creating a transient condition in which tunneling through the internal barrier is maximally enhanced.

This enhanced tunneling prior to the field maximum is also found across the EI regime, independent of the exact geometry (see Supplemental Material for more details). 
This behavior contrasts with standard Keldysh tunneling, where the barrier width decreases monotonically with increasing field strength, such that the ionization rate is maximized at the peak of the field. 
Within the SFA, earlier ionization times map to larger final momenta \cite{Faisal1973_Multiple, Reiss1980_Effect, Lewenstein1994_Theory}, so systems that preferentially ionize before the peak of the field are expected to have a wider distribution of electron momenta, and a distribution that does not peak at zero momentum. 
To evaluate the effect of the molecular potential on this mapping, we use the WKB rate at $E_0=0.17$ a.u. to sample ionization times and tunnel-exit positions and propagate the outgoing electrons using a one-dimensional semiclassical trajectory model. Fig.~\ref{fig:MEP_fig}c compares the resulting final momentum distribution with the SFA prediction for the same ionization-time distribution. Including the molecular potential shifts the predicted momentum offset toward zero, demonstrating that Coulomb interactions substantially reduce the final electron momentum relative to the SFA prediction. Details of the CTMC calculation are given in the Supplemental Material.

The non-Gaussian electron momentum distributions in Fig.~\ref{fig:covariance_1d} are consistent with ionization away from the peak of the field and thus sample a broader range of longitudinal momenta. Additionally, the observation that the relative enhancement of the ionization rate does not increase with probe field strength supports the existence of a critical field strength at which the enhancement is maximized. 
The simple SFA mapping does not quantitatively reproduce the measured momentum offset. At a peak field strength of 0.17 a.u., the maximum WKB rate at an instantaneous field of approximately 0.094 a.u. maps to an SFA momentum offset of approximately 2.5 a.u., compared with the experimental value of approximately 0.7 a.u. The semiclassical trajectory calculation in Fig.~\ref{fig:MEP_fig}c demonstrates that interaction with the molecular potential substantially reduces this offset.

An additional important difference is that the calculations use a continuous-wave (CW) field with a fixed peak amplitude, whereas the experiment uses a few-cycle pulse. The EI rate is maximized when the instantaneous field reaches approximately 0.094 a.u.; it does not require the field to reach the nominal experimental peak of 0.17 a.u. In a few-cycle pulse, the envelope and carrier-envelope phase produce half cycles with different local peak amplitudes. The critical field can therefore be reached near the maximum of a lower-amplitude half cycle, where the vector potential and corresponding final drift momentum are substantially smaller than in the $E_0=0.17$ a.u. CW calculation. Because the experimental pulses are not CEP-stabilized, the measured momentum distribution additionally averages over these different sub-cycle field conditions. The remaining quantitative difference may also reflect limitations of the one-dimensional treatment, as well as other effects not included in the present calculations. Further discussion is given in the Supplemental Material. 
However, the general picture is robust despite these differences. As long as the field-free ionization threshold lies above the internal barrier, and at high fields the system can tunnel through an internal barrier formed by the downhill deuteron, an intermediate field strength must exist where the barrier is transiently narrowest, allowing tunneling over a broader range of times within the optical cycle.

The measured EI electron momentum distributions directly probe the dynamics of the electrons emitted in EI and complete the picture of strong-field EI in simple molecules.
They show that field-induced barrier suppression can maximize tunneling at an intermediate critical field rather than at the peak field in the optical cycle. This in turn generates broader, non-Gaussian electron momentum distributions, consistent with our observations. 
This highlights the importance of the time-dependent structure of the tunneling barrier in determining not only the ionization probability but also properties of photoemission in the strong-field regime.


These observed effects may have broader implications for strong-field light-matter interactions. Because the timing of tunnel ionization determines the subsequent electron trajectories in the laser field, controlling the tunneling dynamics through molecular structure could provide a route to modifying the distribution of high-harmonic emission within the plateau region \cite{Corkum1993_Plasmaa, Schafer1993_threshold, Krause1992_Highorder}. This work motivates improved theoretical descriptions that capture the coupled effects of Stark shifts and evolving molecular geometry to drive resonant effects in the tunneling dynamics of the system. It also points toward a new regime in which molecular structure can be used to actively shape strong-field electron dynamics, opening the possibility of engineering light-driven processes at the level of the tunneling step itself.

\begin{acknowledgments}
We thank M. Ivanov for useful discussions. 
E.W., C.C., I.G., A.J.H., A.M.G., H.M., and P.H.B. were supported by the National Science Foundation. 
A.J.H. was additionally supported under a Stanford Graduate Fellowship as the 2019 Albion Walter Hewlett Fellow. A.M.G. was additionally supported by an NSF Graduate Research Fellowship. 
The data that support the findings of this work are openly available \cite{Weckwerth2026_Data}.
\end{acknowledgments}

\clearpage
\onecolumngrid
\renewcommand{\thepage}{S\arabic{page}}
\setcounter{page}{1}
\renewcommand{\thesection}{S\arabic{section}}
\renewcommand{\thesubsection}{S\arabic{section}.\arabic{subsection}}
\renewcommand{\thesubsubsection}{S\arabic{section}.\arabic{subsection}.\arabic{subsubsection}}
\setcounter{section}{0}
\renewcommand{\thefigure}{S\arabic{figure}}
\setcounter{figure}{0}
\renewcommand{\theequation}{S\arabic{equation}}
\setcounter{equation}{0}
\setstretch{1.25}

\makeatletter
\let\title\SIrestore@title
\let\author\SIrestore@author
\let\affiliation\SIrestore@affiliation
\let\thanks\SIrestore@thanks
\let\date\SIrestore@date
\let\maketitle\SIrestore@maketitle
\let\and\SIrestore@and
\setcounter{affil}{0}
\makeatother

\title{\textbf{Supplemental Materials} for Electron spectra from strong-field enhanced ionization in heavy water} 

\author{Eleanor Weckwerth} 
\address{\PULSE}
\address{\StanfordPhysics}

\author{Chuan Cheng}
\address{\PULSE}
\address{\StanfordPhysics}

\author{Ian Gabalski}
\address{\PULSE}
\address{\StanfordAP}
\address{\Davis}

\author{Andrew~J.~Howard}
\address{\PULSE}
\address{\StanfordAP}
\address{\Davis}

\author{Mathew~Britton}
\address{\PULSE}
\address{\SLAC}

\author{Aaron~M.~Ghrist}
\address{\PULSE}
\address{\StanfordAP}

\author{Haoran Ma}
\address{\PULSE}
\address{\StanfordPhysics}

\author{Salma A. Mohideen}
\address{\PULSE}
\address{\StanfordPhysics}

\author{Philip H. Bucksbaum}
\address{\PULSE}
\address{\StanfordPhysics}
\address{\StanfordAP}
\thanks{Corresponding author: phbuck@stanford.edu}

\date{\today}

\maketitle 

\onecolumngrid
\setstretch{1.25}

\section{Supplemental Experimental Methods}

The few-cycle pulse pairs used in the experiment were generated starting from 40-fs 800-nm Ti:sapphire pulses at 1-kHz repetition rate, which were then spectrally broadened in a 2.5-m long, 500-$\mu$m inner diameter stretched hollow core fiber (HCF) differentially pumped with gaseous Ar to a maximum pressure of approximately $6$~psig \cite{SI-Robinson2006_generation}. The broadened pulses were recompressed to 6-fs duration using chirped mirrors and characterized using a second-harmonic generation dispersion scan with BK7 wedges \cite{SI-Miranda2012_Simultaneous}.  After recompression, the beam passed through a vertical polarizer before being split in a Mach-Zehnder interferometer to produce two pulses with variable delay. Each arm of the interferometer contained a polarizer and a broadband half-wave plate for independent power control, resulting in a pair of cross-polarized pulses with tunable delay and intensity. The pump-probe delay was determined with sub-fs precision from the spectral interference of the combined output beam, measured using a beam sampler and a spectrometer.


The pulse pairs were focused by a $f =$ 10~cm concave mirror contained in an ultra-high vacuum chamber at a base pressure of $2\times10^{-9}$~Torr. The pulse intensities were varied across datasets (see Sec.~S2), with most measurements performed near a peak intensity of $1.0\times10^{15}$~W/cm$^2$, corresponding to a peak electric field strength of 0.17~a.u. 
The absolute peak intensities were calibrated from the measured Ar$^+$/Ar$^{2+}$ ion-yield ratio using the procedure described in Refs.~\cite{SI-Wallace2016_Precise,SI-Weckwerth2025_Optical}. We estimate the absolute intensity calibration to be accurate to within approximately 20\%, corresponding to an uncertainty of approximately 10\% in the peak electric field strength. Relative intensities within each scan were determined from direct power measurements and are accurate to within approximately 3\% (approximately 1.5\% in field strength). 
The chamber was backfilled with gaseous \ddo to a pressure of $8\times10^{-9}$~Torr, yielding a combined ion count rate of $\approx\!1$-$2$ ions per shot. The charged fragments were detected using a double-sided velocity map imaging (VMI) spectrometer and a Timepix3 camera (TPX3CAM) \cite{SI-Eppink1997_Velocitya,SI-Chandler1987_Twodimensional,SI-Zhao2017_Coincidencec,SI-Zhao2017_Coincidence,SI-Gabalski2026_Fasta}. 
The \ddocoin triple ionization events and \dodcoin double ionization events are identified using the same methods described in detail in the Supplemental Materials for Ref.~\cite{SI-Howard2025_Isotopeselective}. 

\section{Ionization rate enhancement for various intensities}

Four datasets were collected with identical conditions except for the pump and probe field strengths and the delay range. In the first dataset, the pump and probe fields were equal at 0.129 a.u., and interpulse delays from 10-80 fs were recorded. In the remaining datasets, the pump field was fixed at 0.155 a.u., while the probe field was varied between 0.155 a.u., 0.170 a.u., and 0.186 a.u., with delays spanning 10-50 fs.

\begin{figure}[htbp]
	\centering
	\includegraphics[width=0.95\linewidth]{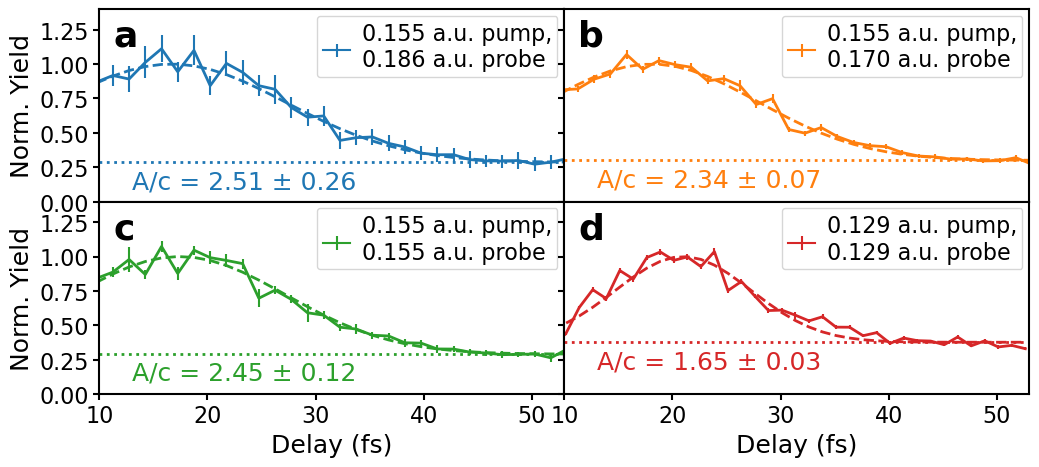}
	\caption{Comparison of the relative amount of enhancement for the \ddocoin channel at four different sets of pump/probe intensities. Data are fit to a Gaussian with a constant offset and the degree of enhancement is quantified by the ratio of the Gaussian amplitude $A$ to the vertical offset $c$, or the rate at the peak enhancement versus the rate for far separated pulses. Each trace is normalized to the maximum yield of the fitted Gaussian for the enhancement. 
    }
	\label{fig:enhance_compare}
\end{figure}

The relative enhancement in the \ddocoin channel, defined with respect to the constant ionization rate expected for two well-separated pulses, is shown in Fig.~\ref{fig:enhance_compare}. Each curve is fit to a Gaussian with a constant offset, and the enhancement is quantified by the ratio $A/c$, corresponding to the peak rate relative to the offset. The figure shows that for the datasets with fixed pump field (panels a-c), the magnitude of the enhancement is effectively independent of the probe field strength over this range. This behavior is consistent with the critical-field picture discussed in the main text: increasing the nominal peak field primarily changes when within the pulse the optimal field for EI is reached rather than substantially increasing the relative EI yield.

However, the dataset in panel d, taken at the lower pump field of 0.129 a.u., exhibits a reduced enhancement. 
The pump pulse prepares a distribution of \ddox$^{2+}$ states and nuclear configurations, only a subset of which evolve along the slingshot pathway and reach the critical geometry for EI near 20 fs. Changing the pump field strength changes the relative population of these pathways and therefore the fraction of the dication ensemble that contributes to the EI peak relative to the delay-independent triple ionization yield. Thus, the reduced enhancement in panel d reflects a change in the dication ensemble prepared by the pump rather than a change in the underlying EI mechanism.

The dataset at a probe field strength of 0.186 a.u. does not have sufficient statistics to extract a reliable delay-dependent electron $p_x$ spectrum in covariance with the \ddocoin channel, and is therefore not included in the electron momentum comparison in Fig.~2b of the main text.  The remaining three datasets are shown there, including two with a pump field of 0.155 a.u. and one with 0.129 a.u. This difference in pump field is not expected to affect the electron momentum distributions in the EI-selected \ddocoin channel. In all cases, the selected events correspond to dications created by the pump pulse and subsequently ionized to the trication by the probe pulse. The observation of a consistent enhancement near 20~fs across all datasets indicates that the pump pulse populates trajectories that evolve along the same ``slingshot'' pathway of coupled stretching and unbending. While the relative population of these trajectories depends on the pump field strength, the dynamics of the selected events, and thus the resulting electron spectra, are governed by the same underlying EI mechanism. 

\section{Pump pulse electron contribution}

\begin{figure}[htbp]
	\centering
	\includegraphics[width=0.7\linewidth]{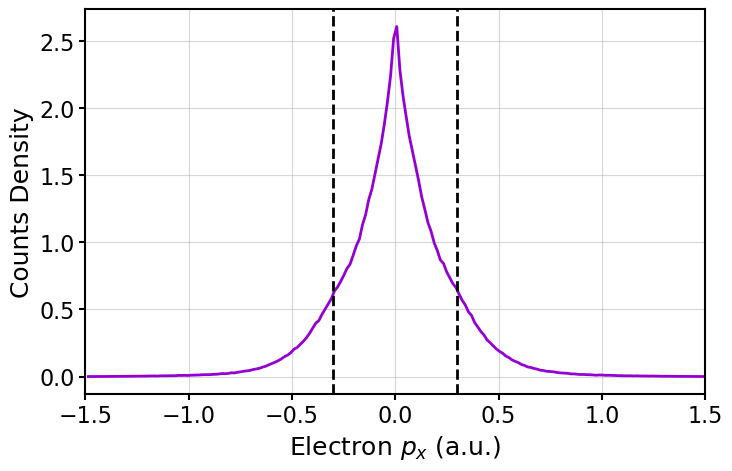}
	\caption{Electron $p_x$ distribution for all electrons detected with the pump pulse alone (field strength 0.155 a.u.), measured by periodically blocking the probe pulse during delay scans. The distribution is strongly localized near zero momentum, with approximately 80\% of detected electrons falling within $|p_x| < 0.3$~a.u., as indicated by the dashed vertical lines.
    }
	\label{fig:pump_alone}
\end{figure}

The electron $p_x$ distributions presented in Fig.~2 of the main text include all electrons detected in covariance with the selected ion channels. For the triple ionization channels, three electrons are produced per event; however, in the EI regime two of these originate from the pump pulse, which prepares the dication that evolves along the ``slingshot'' trajectory. These pump electrons do not significantly affect the distributions shown in Fig.~2 because the pump pulse is polarized along the spectrometer time-of-flight axis, resulting in electrons that are strongly localized near $p_x = 0$ (and near $p_y = 0$). 
This is demonstrated in Fig.~\ref{fig:pump_alone}, which shows the electron momentum distribution obtained with the pump pulse alone. 
Because the pump pulse is polarized along the spectrometer time-of-flight axis ($\hat{z}$), its photoelectron momentum distribution is cylindrically symmetric about this axis. Consequently, the measured two-dimensional projection onto the detector $x$-$y$ plane is azimuthally symmetric, and only the $p_x$ distribution is shown in Fig.~\ref{fig:pump_alone}. 

These data were collected simultaneously with the pump-probe measurements by periodically blocking the probe pulse during delay scans. The distribution is sharply peaked at low momentum, with approximately 80\% of the detected electrons falling within $|p_x| < 0.3$~a.u. This strong localization confines the pump electron contribution to a narrow region near the center of the combined spectra.
The same single pulse measurements were also used to quantify direct triple ionization by the pump (or probe) pulse. The \ddocoin coincidence rate with one pulse blocked is less than 2\% of the rate measured at late pump-probe delays when the two pulses are well separated, confirming that direct trication production by a single pulse is negligible under the conditions used here.

The true distribution of pump electrons is likely even more concentrated than indicated in Fig.~\ref{fig:pump_alone}. The detection efficiency of the MCP is reduced near the center of the detector, where the flux of parent ions is highest. As a result, electrons with small transverse momentum are undercounted, and the measured distribution underestimates the degree of central localization. While this effect is difficult to quanitfy precisely, since it depends on delay, intensity, and ion species, the net result is that the pump electron contribution is even more strongly confined to low $p_x$ than observed. 

A further suppression of the pump electron contribution arises from event overlap on the detector. When two electrons arrive within a small spatial separation (on the order of the detector resolution), they cannot be distinguished as separate hits. Because pump electrons are concentrated near the center of the detector and are present in essentially every shot (both from the selected ionization event and from background ioniziation), they have a higher probability of overlapping and being undercounted. In contrast, probe electrons are distributed more broadly across the detector and are therefore much less likely to overlap. This effect further biases the combined spectra toward probe electrons.

Finally, the pump electron contribution is similar across all ion channels considered here. Any channel-dependent differences in the measured electron spectra therefore arise from variations in the probe-electron distribution. For these reasons, we treat the combined electron spectra as being dominated by probe electrons. 
To further minimize any effects from small contributions from pump electrons, the central region ($|p_x| < 0.2$~a.u.) is excluded from the combined Gaussian fits described in the main text.

\section{Sensitivity to momentum exclusion window}

\begin{figure}[htbp]
	\centering
	\includegraphics[width=0.9\linewidth]{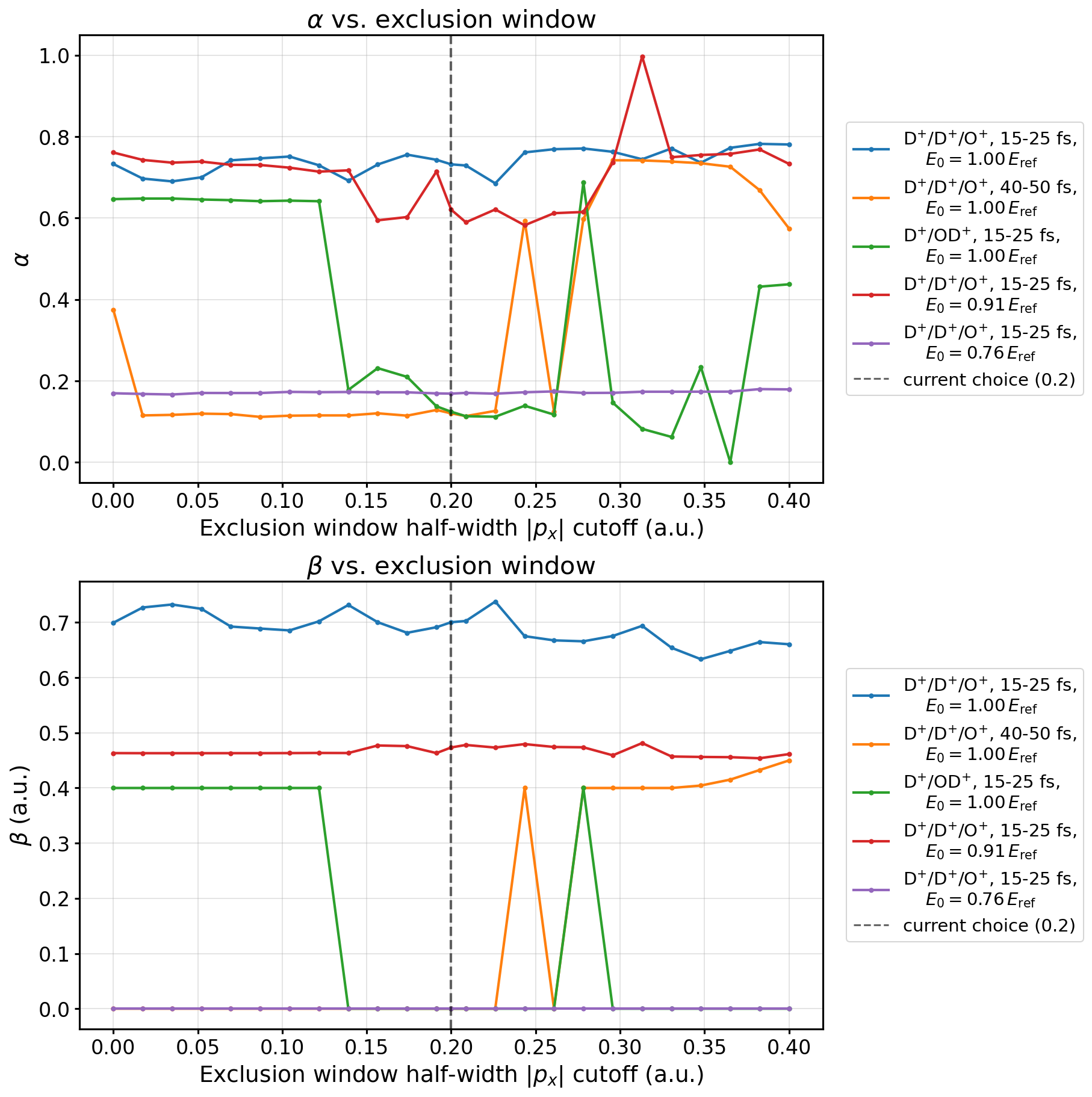}
	\caption{Sensitivity of the Gaussian fit parameters to the central momentum exclusion window. Extracted \textbf{(a)} relative offset contribution $\alpha$ and \textbf{(b)} primary offset $\beta$ as the width of the excluded region $|p_x|<p_{\mathrm{cut}}$ is varied from 0 to 0.4~a.u. The spectra and line colors correspond to those shown in Fig.~2 of the main text. The vertical dashed line indicates the value $p_{\mathrm{cut}}=0.2$~a.u. used for the reported fits.
    }
	\label{fig:window_sens}
\end{figure}

To evaluate the sensitivity of the fitted parameters to the size of the excluded central region, the fits to Eq.~(1) in the main text, shown in Fig.~2 of the main text, were repeated while varying the exclusion window half-width from 0 to 0.4~a.u. Fig.~\ref{fig:window_sens} shows the resulting values of $\alpha$ and $\beta$ for the spectra analyzed in Fig.~2 of the main text. The vertical dashed line indicates the value $|p_x|<0.2$~a.u. used in the main text.

For the EI \ddocoin spectra, the qualitative behavior of the fitted parameters is stable over a broad range of exclusion-window sizes: the highest-field EI spectrum retains a substantial nonzero momentum offset, while the offset decreases as the probe field is reduced. The non-EI spectra show larger fluctuations in the fitted parameters for some exclusion windows. Because these distributions contain a larger fraction of their yield near $p_x=0$, increasing the exclusion window removes a greater fraction of the spectral region that constrains the centered component of the fit, making the relative centered and offset contributions less well determined. The chosen cutoff of 0.2~a.u. lies within a region where the fitted parameters retain the qualitative distinction between the EI and non-EI spectra.

\section{Derivation of Gaussian momentum distribution in strong-field ionization}

Ionization of molecules in strong sub-picosecond infrared laser fields can be described as a Keldysh tunneling process through a barrier formed by the Coulomb potential and the oscillating linear potential of the laser field. In the adiabatic limit, the tunneling rate is proportional to the negative exponential of the reciprocal of the field \cite{SI-Keldysh1964_Ionization,SI-Ammosov1986_Tunnelb,SI-Perelomov1966_Ionization}:
\begin{equation}
W(t) \propto
\exp{\left(-\frac{3}{2}\frac{I_p^{-3/2}}{|E(t)|}\right)},
\end{equation}
where $I_p$ is the bound state ionization potential, and both $I_p$ and $E(t)$ are in atomic units.
The proportionality factor depends on details of the charge distribution and the level of sophistication of the model; but the exponential field dependence is nearly universal across all models for coulomb-bound systems. (See \cite{SI-Ditmire2025_Strong} for a recent review.)

For a linearly  polarized laser field we may expand $E(t)$ in a Taylor series near its maximum at $t=0$:
\begin{equation}
E(t)=E_0\cos(\omega t)
 \approx E_0(1 - \omega^2 t^2/2),
\end{equation}
which leads to a Gaussian time dependence of the ionization rate during the optical cycle:
\begin{equation}
W(t) \propto
\exp{\left(-\frac{3}{4}\frac{I_p^{-3/2}\omega^2t^2}{E_0}\right)}.
\end{equation}

The Strong Field Approximation  (SFA) assumes the initial longitudinal momentum $[p_\parallel]_{init}$ of the electron is zero as it exits the tunnel barrier, and it evolves solely under the influence of the laser field
\cite{SI-vandenHeuvell_Limiting,SI-Corkum1993_Plasmaa}.
It can be described as a Volkov solution of the free-electron Hamiltonian \cite{SI-Keldysh1964_Ionization,SI-Volkov1935_Ueber}.
\begin{equation}
H_{free}=
\frac{|\textbf{p} - \textbf{A}(t)|^2}{2}
\end{equation}
where $\textbf{A}$ is the classical laser vector potential.
The cycle-averaged kinetic energy is conserved, so  
ioinzation at $t=t_0$ results in an electron with final longitudinal momentum 
\begin{equation}
p_{\parallel} = -\textbf{A}(t_0).
\end{equation}

Expanding the time-dependent vector  potential about $t=0$ and keeping the lowest order (linear) dependence
\begin{equation}
\textbf{A}(t) =
-\frac{E_0}{\omega} \sin{\omega t}
\approx
-E_0 t
\end{equation}
maps the Gaussian time dependence of the ionization rate onto momentum space. This yields a Gaussian photoelectron momentum distribution,
\begin{equation}
\textrm{PMD}(p_\parallel) \propto
\exp{\left(-\frac{3}{4}\frac{I_p^{-3/2}\omega^2p_\parallel^2}{E_0^3}\right)}.
\end{equation}

\section{Details of WKB calculations}

The molecular electrostatic potential (MEP) for the stretched linear \ddox$^{3+}$ geometry shown in Fig.~3a of the main text was computed using restricted open-shell Hartree-Fock (ROHF) theory in GAMESS~\cite{SI-Barca2020_Recent} with a 6-31G Gaussian basis set. The MEP was interpolated onto a dense spatial grid, and one-dimensional lineouts were extracted along the laser polarization direction for the selected molecular orientation. An external time-dependent electric field with peak amplitude $E_0 = 0.17$~a.u. was applied to the MEP along the tunneling coordinate to yield an effective potential.

Field-dependent ionization potentials for \ddox$^{2+}$ were obtained from multiconfigurational self-consistent field (MCSCF) calculations using correlation-consistent Gaussian basis sets, and interpolated as a function of instantaneous field strength to account for Stark shifts. For the rates shown in Fig.~3b of the main text, at each time step, classical turning points were identified along the downhill direction of the field, and the tunneling probability was evaluated within a one-dimensional WKB framework by integrating across the classically forbidden region between the inner and outer turning points
\begin{equation}
\Gamma(t) \propto \exp\!\left[-2 \int_{x_{\mathrm{in}}}^{x_{\mathrm{out}}} \sqrt{2\left(V_{\mathrm{eff}}(x,t) - E_{\mathrm{IP}}(t)\right)}\,dx \right],
\end{equation}
where $x_{\mathrm{in}}$ and $x_{\mathrm{out}}$ are the classical turning points. The applied field is taken to be spatially uniform and sinusoidal in time, without inclusion of a pulse envelope, and the tunneling rate is evaluated in the quasi-static (instantaneous field) approximation.

\section{Semiclassical trajectory calculations}

To estimate the influence of the molecular potential on the final electron momentum, we performed one-dimensional semiclassical trajectory calculations for the same linear stretched geometry used in Fig.~3 of the main text ($r_{\mathrm{OD}}=2.50$~\AA, $\theta_{\mathrm{DOD}}$=180$^{\circ}$). For the calculation shown in Fig.~3c, $2\times10^4$ ionization times were sampled from the WKB tunneling-rate distribution at a peak field strength $E_0=0.17$~a.u. The corresponding tunnel-exit position for each trajectory was taken from the outer classical turning point of the WKB barrier, and each electron was initialized with zero longitudinal velocity.

The trajectories were propagated in the sinusoidal laser field and the one-dimensional field-free molecular electrostatic potential. Outside small regions surrounding the atomic centers, the molecular force was obtained directly from the derivative of the calculated MEP. Within a splice radius of 0.10~\AA\ around each atomic site, the potential was replaced by a soft-core Coulomb form with softening parameter a=0.08~\AA, with the effective charge chosen to match the calculated MEP at the splice boundary. Equivalent ionization in the opposite field direction was included to produce the symmetric momentum distribution.

For comparison, the SFA momentum for the same set of sampled ionization times was calculated directly from the laser vector potential. Trajectories that returned to the oxygen core region were excluded from the CTMC distribution shown in Fig.~3c of the main text. The inclusion of the molecular potential shifts the dominant finite-momentum features toward zero relative to the SFA distribution, showing that Coulomb interaction with the molecular ion provides a significant correction to the simple SFA phase-to-momentum mapping.

\section{Geometry and field strength dependence of peak phase}

The details of the tunneling rate versus phase calculation depend strongly on the molecular geometry used in the model. In the main text, the calculations are performed using a symmetric O-D bond length of 2.50~\AA\ to match the enhancement bond length determined in Ref.~\cite{SI-Howard2025_Isotopeselective}. However, the \ddo enhanced ionization (EI) regime is expected to occur over a range of molecular geometries.

To explore how the bond length within the EI regime affects our calculations, we carried out the same MEP and IP calculations for O-D bond lengths between 2.2 and 2.55~\AA, using a peak field strength of 0.17~a.u. and a DOD bond angle of 180$^{\circ}$. One important difference that appears is that for longer bond lengths, both the internal tunneling barrier (between the oxygen and the downhill deuterium) and the external barrier are present simultaneously over part of the field cycle, as shown in Fig.~3a in the main text. In this calculation, for bond lengths shorter than approximately 2.4~\AA, the external barrier drops below the instantaneous ionization potential before the internal barrier rises above it. As a result, the calculated tunneling rates for these shorter bond lengths become discontinuous in phase. In this region, we model the tunneling rate as zero even though the calculation suggests it will result in over-the-barrier ionization. The precise treatment of this regime is uncertain, and the details of the calculation, such as the localization of the ionizing orbital, are likely to play an important role. Here, we adopt a simplified approach intended to capture the overall trend.

\begin{figure}[htbp]
	\centering
	\includegraphics[width=0.95\linewidth]{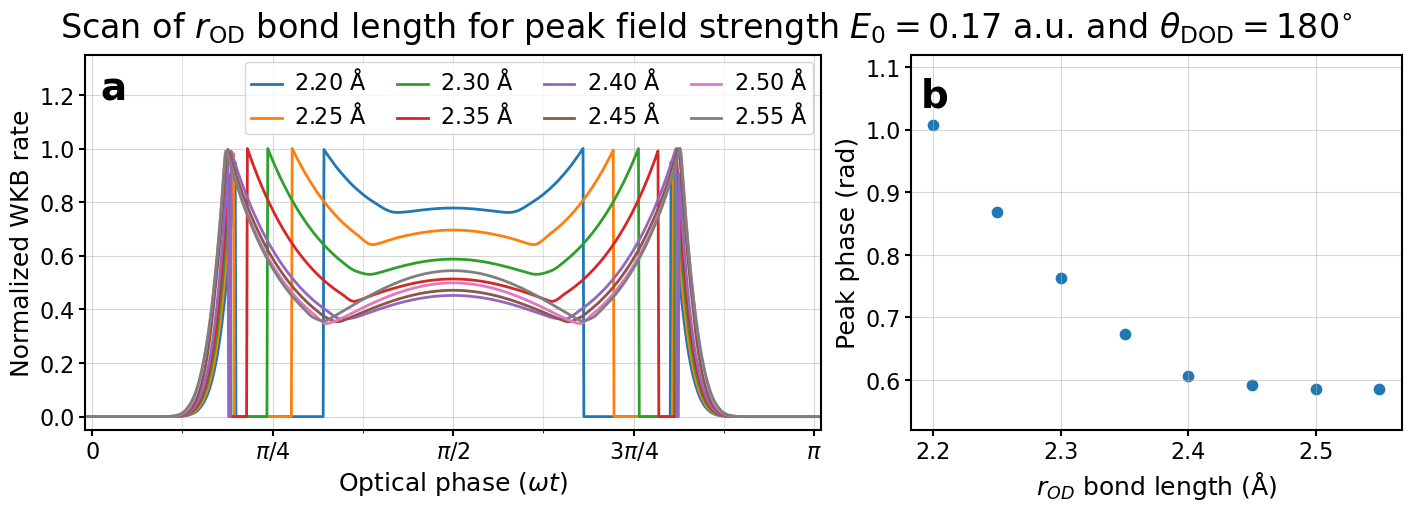}
	\caption{Dependence of the WKB tunneling rate on O-D bond length for a peak field strength of 0.17~a.u. and $\theta_{\mathrm{DOD}} = 180^{\circ}$.
    \textbf{(a)}~Normalized tunneling rate as a function of optical phase for bond lengths between 2.2 and 2.55~\AA. For shorter bond lengths, the tunneling rate becomes discontinuous in phase due to the absence of a well-defined tunneling barrier over part of the field cycle, and is set to zero in this region.
    \textbf{(b)}~Phase at which the tunneling rate is maximized as a function of bond length.
    }
	\label{fig:bondlength}
\end{figure}

The resulting tunneling rate curves are shown in Fig.~\ref{fig:bondlength}a, and the phase corresponding to the maximum tunneling rate for each bond length is shown in Fig.~\ref{fig:bondlength}b. As the bond length increases, the peak phase shifts to earlier times in the field cycle (farther from the field maximum). This trend continues until approximately 2.4~\AA, beyond which both barriers are present simultaneously. In this regime, the shift in peak phase becomes more gradual and approaches an asymptotic value near 0.58~rad, as shown in the main text calculations.

The phase where the tunneling rate is maximized also depends on the peak field strength, since the tunneling probability is determined by the instantaneous field. For the geometry used in the main text (2.50~\AA, $\theta_{\mathrm{DOD}} = 180^{\circ}$), the maximum tunneling rate occurs at an instantaneous field strength of 0.094~a.u. As a result, the corresponding phase varies depending on the peak field amplitude. Since the phase is really what determines the momentum offset, this turns out to be an important distinction.

\begin{figure}[htbp]
	\centering
	\includegraphics[width=0.95\linewidth]{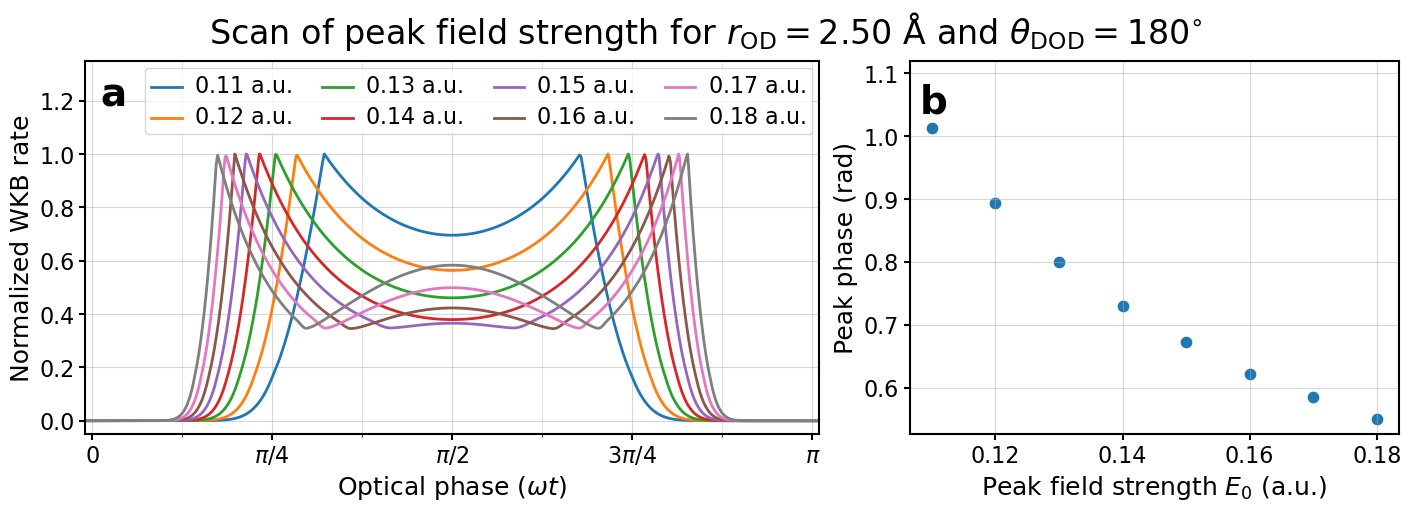}
	\caption{Dependence of the WKB tunneling rate on peak field strength for a fixed geometry (O-D bond length of $2.50$~\AA, $\theta_{\mathrm{DOD}} $ of $180^{\circ}$).
    \textbf{(a)}~Normalized tunneling rate as a function of optical phase for peak field strengths between 0.11 and 0.18~a.u. Increasing the peak field shifts the phase at which the tunneling rate is maximized.
    \textbf{(b)}~Phase of maximum tunneling rate as a function of peak field strength.
    }
	\label{fig:fieldstrength}
\end{figure}

This distinction is particularly important because the calculations in the main text use a continuous-wave (CW) field without an envelope, while the experiment uses a 6-fs few-cycle pulse. In the experimental pulse, the envelope and CEP cause successive half cycles to have different local peak amplitudes. EI can therefore occur when a lower-amplitude half cycle reaches the critical field strength, placing the preferred ionization time closer to that half-cycle's field maximum. Since the vector potential approaches zero at a field maximum, this results in a smaller final momentum than the $E_0=0.17$ a.u. CW mapping. The CW calculation therefore provides a qualitative connection between the critical-field tunneling mechanism and the momentum distribution, rather than a quantitative prediction of the experimental offset. To illustrate the sensitivity of the ionization phase to the local peak field strength, tunneling-rate curves for the same geometry as in the main text, but with varying peak field strengths, are shown in Fig.~\ref{fig:fieldstrength}. Because the experimental pulses are not CEP-stabilized, the measured momentum distribution represents an average over different CEPs and therefore over different local half-cycle peak fields and corresponding ionization phases. This averaging is expected to broaden the mapping between ionization phase and final momentum relative to the single-phase CW calculation.

\end{document}